\documentclass[aps,prl,reprint,superscriptaddress]{revtex4-1}
\usepackage{graphicx}
\usepackage{dcolumn}
\usepackage{bm}
\usepackage{amsmath}

\usepackage{amsfonts}
\usepackage{textcomp} 

\usepackage{verbatim}
\usepackage {ulem}

\usepackage{color}
\usepackage{array}
\usepackage{makecell}

\newcolumntype{C}{>{\centering\arraybackslash$}m{2cm}<{$}}

\begin{document}

\title{Evolution of the interfacial perpendicular magnetic anisotropy constant of the Co$_2$FeAl/MgO interface upon annealing}

\author{A. Conca}

\email{conca@physik.uni-kl.de}

\affiliation{Fachbereich Physik and Landesforschungszentrum OPTIMAS, Technische Universit\"at
Kaiserslautern, 67663 Kaiserslautern, Germany}

\author{A.~Niesen}

\author{G.~Reiss}

\affiliation{Center for Spinelectronic Materials and Devices, Physics Department, Bielefeld University, 100131 Bielefeld, Germany}

\author{B.~Hillebrands}

\affiliation{Fachbereich Physik and Landesforschungszentrum OPTIMAS, Technische Universit\"at
Kaiserslautern, 67663 Kaiserslautern, Germany}

\date{\today}

\begin{abstract}
We investigate   a series of films with different thickness of the Heusler alloy Co$_2$FeAl in order to study the effect of annealing on the interface with a MgO layer and on the bulk magnetic properties.  Our results reveal that while the perpendicular interface anisotropy constant $K^{\perp}_{\rm S}$ is zero for the as-deposited samples, its value increases with annealing up to a value of  $1.14\, \pm \,0.07$~mJ/m$^2$ for the series annealed at 320$^{\rm o}$C  and of $2.01\, \pm \,0.7$~mJ/m$^2$ for the 450$^{\rm o}$C annealed series owing to a strong modification of the interface during the thermal treatment. This large value ensures a stabilization of a perpendicular magnetization orientation for a thickness below 1.7~nm. The data additionally shows that the in-plane biaxial anisotropy constant has a different evolution with thickness in as-deposited and annealed systems. The Gilbert damping parameter $\alpha$ shows minima  for all series for a thickness of 40~nm and an absolute minimum value of $2.8\pm0.1\times10^{-3}$. The thickness dependence is explained in terms of an inhomogeneous magnetization state generated by the interplay between the different anisotropies of the system and by the crystalline disorder. 

\end{abstract}

\maketitle

\section{Introduction}

In order to achieve efficient spin torque switching, materials with a certain set of properties are required. These properties are a combination of low damping and low magnetization, together with the presence of a robust perpendicular magnetic anisotropy (PMA). Additionally, these materials should have a high spin polarisation and be compatible with standard tunneling barrier materials such as MgO or MgAl$_2$O$_4$. A high Curie temperature is also desirable to guarantee temperature stability. 

In the wide family of the Heusler compounds, some candidates can be found which  fulfill the aforementioned requirements. For instance, large tunneling magnetoresistance (TMR) ratios have been  reported for several compounds \cite{ando-co2mnsi,inomata-cfas,inomata-cfas2,yamamoto-co2mnsi,andy-cfa,drewello-heusler}. Heusler films have been successfully employed in systems with PMA   \cite{niesen-tin,takamura-co2fesi,kamada-cfms,lufbrook-cfms-pma,lufbrook-cmnga-pma}  and show also low damping properties \cite{oogane-cfs-damping}. For the PMA properties of thin Heusler films,  the interface-induced perpendicular anisotropy plays a critical role and its strength is given by the value of the perpendicular interfacial anisotropy constant $K^{\perp}_{\rm S}$. The interfacial properties, and therefore the value of the constant, are strongly modified by the exact  conditions of the annealing treatment for the stack, which is required to improve the crystalline order of the Heusler films \cite{drewello-heusler,cinchetti,conca-ccfa}  and to achieve large TMR values \cite{schebaum}. The alloy Co$_2$FeAl belongs to the materials for which large TMR \cite{wang-cfa-tmr} have been reported, even for textured films on a SiO$_2$ amorphous substrate \cite{wen-cfa}. Low damping \cite{mizukami-cfa-damping,cui-cfa-damping-pma,cui-cfa-damping} and PMA \cite{wen-cfa-pma,cui-cfa-damping-pma} have also been  proven. In this work, we study the evolution with annealing of $K^{\perp}_{\rm S}$ in systems with a MgO interface,  by measuring different thickness series. Since the in-plane anisotropies and the Gilbert damping parameter change  with varying thickness and annealing temperature, also their evolution is reported. The relevance of the study is not limited to Co$_2$FeAl but it is a model for all TMR systems with Co-based Heusler alloys and an interface with a MgO tunneling barrier.

\begin{figure}[b]
    \includegraphics[width=0.8\columnwidth]{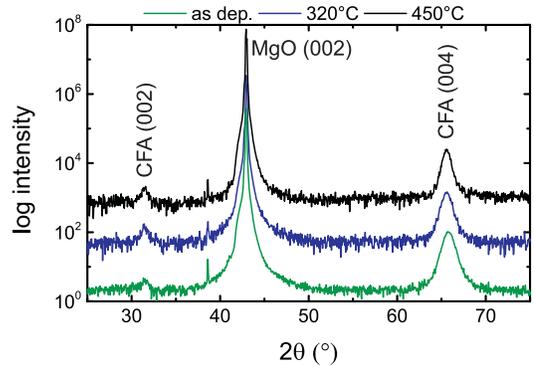}
	  \caption{\label{XRD}(Color online) X-ray diffraction patterns of 20\,nm thin CFA layers as-deposited, annealed at 320$^{\rm o}$C and annealed at 450$^{\rm o}$C. The (002) superlattice and the fundamental (004) peak of the CFA are clearly visible, confirming the partial $B2$ crystalline order.}
\end{figure}

\section{Sample preparation}

Thickness series (7-80~nm) of Co$_2$FeAl (CFA) epitaxial films were prepared and a microstrip-based VNA-FMR setup was used to study their magnetic properties. The dependence of the in-plane anisotropies and the Gilbert damping parameter on the thickness and the determination of the interface perpendicular anisotropy constant K$^{\perp}\rm_S$ for the CFA/MgO interface is presented for as-deposited samples and for two different values of the annealing temperature.

The stack layer structure is  MgO(100)(subs)/ MgO(5)/CFA($d$)/MgO(7)/Ru(2) with $d=7$, 9, 11, 15, 20, 40 and 80~nm.  Rf-sputtering was used for the MgO deposition and dc-sputtering for the rest. The values of the annealing temperature for the two series with thermal treatment are 320$^{\rm o}$C and 450$^{\rm o}$C. The layer stacking is symmetrical around CFA so that a similar interface is expected for both sides. The samples were all deposited at room temperature and  annealed afterwards under vacuum conditions.

\section{X-Ray characterization}

Crystallographic properties of the CFA thin films were determined using x-ray diffraction (XRD) measurements in a Philips X'Pert Pro diffractometer equipped with a Cu anode. The (002) superlattice and the fundamental (004) peak of the CFA can be observed (see Fig.~\ref{XRD}) already for the as-deposited state. In-plane performed $\phi$ scan measurements reveal the absence of the (111) superlattice reflection in these films. Therefore, partial $B2$ crystalline order is verified. 
Epitaxial, 45$^{\rm o}$ rotated growth, relative to the MgO buffer layer, was verified using a $\phi$ scan of the reflection from the (202) planes (not shown here). The epitaxial relationship CFA (001)[100] // MgO(001)[110] was therefore confirmed for these films, i.e. CFA grows with the same  crystalline orientation as the substrate but the unit cell is rotated 45$^{\rm o}$ in plane respect to the MgO unit cell.

\begin{figure}[t]
    \includegraphics[width=0.8\columnwidth]{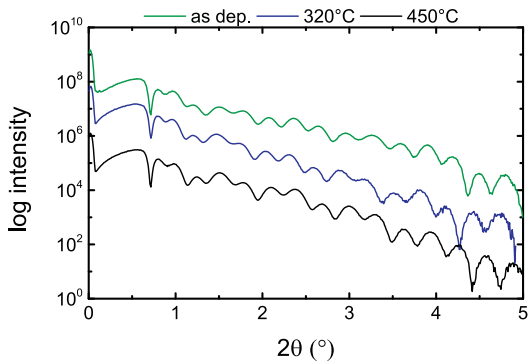}
	  \caption{\label{XRR}(Color online) X-Ray reflectometry data corresponding to samples with a CFA thickness of 20~nm and different annealing temperatures.}
\end{figure}

X-ray reflectometry (XRR) has been performed on the 20~nm thick films and it is shown in Fig.~\ref{XRR}.
The estimation of the RMS value is only possible with a certain uncertainty due to the number of layers
which increases the number of fitting parameters but it is possible to say that it lays around 0.1-0.3~nm for the three samples. In any case, it is evident that the interface is very smooth in all cases and that the annealing is not modifying the roughness properties.

\section{Results and discussion}

From the dependence of $H_{\rm FMR}$ on the resonance frequency $f_{\rm FMR}$, the effective magnetization $M_{\rm eff}$ is extracted using a fit to Kittel's formula \cite{kittel}. 
For a more detailed description of the FMR measurement and analysis procedure please see Ref.~\cite{fept}. $M_{\rm eff}$ is  related to the saturation magnetization of CFA by \cite{liu,nascimento,beaujour} 

\begin{equation} \label{kp}
M_{\rm eff}= M_{\rm s}-H^{\perp}_{K}= M_{\rm s}-\frac{2K^{\perp}_{\rm S}}{\mu_0 M_{\rm s} d}
\end{equation}
where $K^{\perp}_{\rm S}$ is the perpendicular surface (or interfacial) anisotropy constant.

\begin{figure}[t]
    \includegraphics[width=0.8\columnwidth]{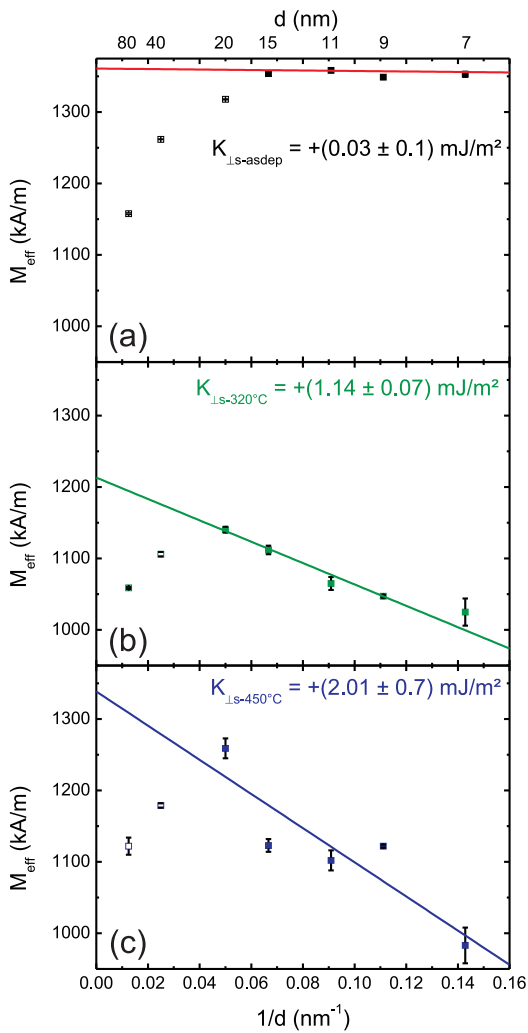}
	  \caption{\label{ks-annealing}(Color online) Dependence of $M_{\rm eff}$ extracted from the Kittel fit on the inverse thickness $1/d$ for three sample series: (a) as-deposited, (b) annealed at 320$^{\rm o}$C, (c) annealed at 450$^{\rm o}$C. The lines are a fit to Eq.~\ref{kp}, the hollow data points were not considered.}
\end{figure}

Fig.~\ref{ks-annealing} shows  the dependence of $M_{\rm eff}$ on $1/d$ for the three CFA series. The lines are a fit to Eq.~\ref{kp}. Let us first discuss the case of the as-deposited series shown in Fig.~\ref{ks-annealing}(a). An almost constant value for $M_{\rm eff}$ is observed for the low thickness range (15-7~nm) where the interface properties should become dominating. The fit gives a value for $K^{\perp}_{\rm S}$ of $0.03\, \pm \,0.1$~mJ/m$^2$ compatible with zero (hollow values  in Fig.~\ref{ks-annealing} not considered for the fit). This implies that it is not possible to obtain a stable perpendicular magnetization orientation for any thickness value based only on the interface effect. However, it has to be  commented that a non-vanishing volume perpendicular anisotropy has also been reported for CFA \cite{wen-cfa-pma} which may indeed stabilize an out-of-plane orientation. Concerning the relative decrease of $M_{\rm eff}$ for large thicknesses, we attribute this to a inhomogeneous magnetization state which is sometimes observed in thick films \cite{chen}. This point will be later commented when analyzing the damping properties.

Figs.~\ref{ks-annealing}(b) and (c) show the evolution of the situation when the annealing step is applied. The interface properties change with the thermal treatment and $K^{\perp}_{\rm S}$ increases to a value  of $1.14\, \pm \,0.07$~mJ/m$^2$ for the 320$^{\rm o}$C case and of $2.07\, \pm \,0.7$~mJ/m$^2$ for 450$^{\rm o}$C. The larger error bar in the later value is due to a larger scattering of values for $M_{\rm eff}$. A recent study of the perpendicular anisotropy properties on CFA thin films has been published where a novel TiN buffer layer is employed \cite{niesen-tin}. In- and out-of-plane hysteresis loops are used to determine the value of $K^{\perp}_{\rm S}$ instead of the FMR measurements used here. However, the largest obtained values for $K^{\perp}_{\rm S}$ are in both cases in accordance with ours ($0.86\, \pm \,0.16$~mJ/m$^2$). For comparison it has to be taken into account that due to the presence of two CFA/MgO interfaces, the values presented here are expected to be a factor of two larger. Both values are then in good agreement. The different annealing temperature range does not allow for a comparison of the evolution of $K^{\perp}_{\rm S}$ with that parameter but a remarkable difference can be found in the as-deposited samples. A comparatively smaller but, contrary to our case, non-zero value is reported. This reveals the role of the TiN buffer layer in improving the interface quality. 

Although it cannot be quantified by XRD, the existence of a certain level of stress in the films cannot be excluded. This stress is changing upon annealing together with the crystalline order at the interface and therefore it is reasonable to admit that it plays a role in the evolution of $K^{\perp}_{\rm S}$. However, it is not possible to separate the contribution to the evolution of the PMA due to these two effects. First principle calculations of $K^{\perp}_{\rm S}$ for stress-free CFA/MgO interfaces \cite{vadapoo} has provided a value for $K^{\perp}_{\rm S}$ of 1.31 mJ/m$^2$ for Co-terminated interfaces while FeAl-termination does induce in-plane orientation. This value is compatible with our results for the 450$^{\rm o}$C case taking into account that our samples have two CFA/MgO interfaces. In any case, our results are more compatible with a Co-termination at the MgO interfaces following this calculation. Other experimental results using XMCD attribute, contrarily to the previous calculation, a PMA contribution to the Fe atoms at the interface \cite{wen2017}. The exact atomic origin of the PMA is then still under discussion and therefore also the actual impact of stress.

As already shown in Fig.~\ref{XRR}, the roughness remains unchanged after the annealing process. The increase  of $K^{\perp}_{\rm S}$ is then due to a more subtle change of the atomic ordering at the immediate interface and is not connected to a roughness modification, or at least not in a large degree.

 By setting $d= \infty$ in Eq.~\ref{kp} it is possible to extract a value for $M_{\rm s}$ of $1140 \pm 30$~kA/m from the linear fit for the as-deposited samples.  This value is larger than the ones reported in  \cite{belme,ortiz} (1000-1030~kA/m) but similar to a FMR study \cite{yadav} on very thick (140~nm) CFA polycrystalline films providing a value of $M_{\rm eff}=1200$~kA/m.

The saturation magnetization $M_{\rm s}$ for TiN buffered CFA, deposited and investigated by the same group, was measured to be $1140 \pm 60$~kA/m, which is in excellent agreement with the value obtained from the FMR data. The saturation magnetization for TiN buffered CFA was obtained using alternating gradient magnetometer (AGM) measurements and verified using vibrating sample magnetometry (VSM) on a 10~nm thin CFA layer \cite{niesen-tin}.

The value of $M_{\rm s}$ also increases upon annealing up to $1213 \pm 8$~kA/m for the 320$^{\rm o}$C series and $1340 \pm 70$~kA/m for the 450$^{\rm o}$C one. This increase can be attributed to an improvement of the crystalline order with annealing.

From the extrapolation of the linear fits to  $M_{\rm eff}=0$ it is possible to extract the thickness at which the interfacial perpendicular anisotropy is able to stabilize an out-of-plane configuration by overcoming the demagnetization field and allowing the magnetic easy axis to be out-of-plane.  This thickness is 1.2~nm and 1.7~nm for 320$^{\rm o}$C and 450$^{\rm o}$C annealing temperature, respectively. The relative difference between both values for the critical thickness is smaller than the relative difference for $K^{\perp}_{\rm S}$ for the respective temperature values. This is explained by the larger $M_{\rm s}$ value for the 450$^{\rm o}$C case for which a larger demagnetizing field must be overcome to achieve PMA.


\begin{figure}[t]
    \includegraphics[width=0.9\columnwidth]{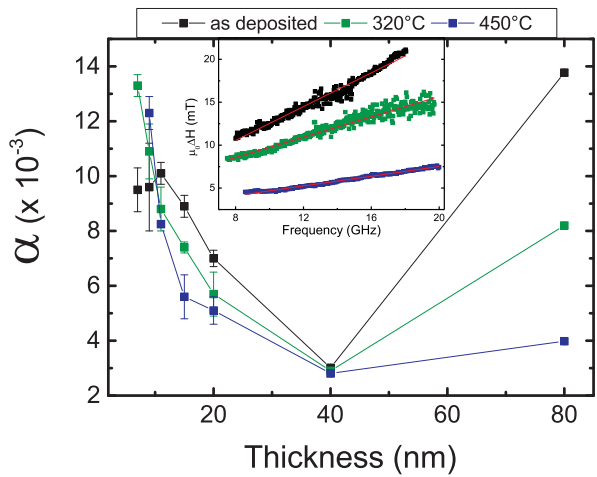}
	  \caption{\label{damping}(Color online) Dependence of the Gilbert damping parameter $\alpha$ on the thickness $d$ for three sample series:  as-deposited, annealed at 320$^{\rm o}$C, and annealed at 450$^{\rm o}$C. The inset shows the dependence of the linewidth $\Delta H$ on the frequency for the 80~nm samples. The lines are a linear fit used to extract the damping parameter $\alpha$. }
\end{figure}

Belmeguenai \textit{et al.}   presented data very similar to the one shown in  Fig.~\ref{ks-annealing}(a) for (110)-ordered  textured films \cite{belme} and for (100)-oriented epitaxial films grown on MgO(100) substrates \cite{belme2}. The annealing temperature is 600$^{\rm o}$C.  The data is given for thickness values not smaller than 10~nm. However, the interpretation of the data is completely opposite to ours, resulting in a negative value $K^{\perp}_{\rm S} = -1.8$~mJ/m$^2$.  The negative value indicates that the interface anisotropy is favoring an in-plane orientation of the magnetization.  
PMA  with Ta/CFA/MgO (or Cr or Ru) systems have been indeed achieved \cite{gabor,wen,wen2} with values of $K^{\perp}_{\rm S} = +0.6$~mJ/m$^2$ for the Ta case,  +1.0~mJ/m$^2$ for Cr and +2.0~mJ/m$^2$ for Ru. This shows how sensitive $K^{\perp}_{\rm S}$ is to the exact growth properties which are modified by the different seed layer. The values reported in this work for both annealed series are very similar to the Cr and Ru buffered systems. The fact that $K^{\perp}_{\rm S}$ vanishes in the as-deposited series shows also how important the annealing step is for adjusting the interface properties.

Figure~\ref{damping} shows the dependence of the Gilbert damping parameter $\alpha$ on the thickness $d$ for the as-deposited samples and the annealed series. The inset shows exemplarily for the 80~nm samples the dependence of the linewidth $\Delta$H on the frequency and the linear fits to obtain $\alpha$. For the three series we observe a minimum in the $\alpha$ value for $d=40$~nm. The smallest value obtained for this series is $\alpha = 2.8\pm0.1\times10^{-3}$. When comparing to the literature it has to be taken into account that the value of $\alpha$ is very sensitive to the growth conditions and to the annealing temperature. Therefore the scatter of values is large. The smallest reported value \cite{mizukami} is around 1$\times 10^{-3}$ but for films annealed at 600$^{\rm o}$C. The damping increases when the annealing temperature is lower, up to values similar to the ones reported here at $\sim$450$^{\rm o}$C.

    The reasons for the increased damping  are different for the thicker and the thinner films. Concerning the large damping value for the 80~nm samples, it is a common behavior in soft magnetic thin films  that the damping increases strongly with thickness starting at a certain value. An example of this can be seen for NiFe in the literature \cite{chen}. In this case the damping of the films strongly increases starting at  $d =90$~nm. The reason for that is a non-homogeneous magnetization state for thicker films which open new loss channels in addition to two-magnon scattering responsible for Gilbert-like behavior in in-plane magnetized films. Nevertheless, the value of $\alpha$  decreases with the annealing temperature pointing to a overall improvement of the uniformity of the film and of the crystalline order.

For the thinner samples down to 11~nm we also observe a reduction of $\alpha$ upon annealing, however this situation is inverted for $d<11$~nm and provides a hint to one of the posible reasons for the increase of damping with decreasing thickness. When the thickness is reduced and the effect of the interface anisotropy is becoming larger the magnetization state is becoming more inhomogeneous due to the counterplay between the demagnetization field and the anisotropy field. However, this is not the only reason explaining the $\alpha$ increase since this is also observable in the as-deposited sample series where $K^{\perp}_{\rm S} \approx 0$, although to a lower degree, and additional effects, e.g. due to roughness, play also a role.

A comment has to be done concerning the exact meaning of the concept of inhomogeneous magnetization used to describe our films. In an ideal thin film with smooth interfaces and in the case of $K^{\perp}_{\rm S}=0$, the demagnetizing field due to the shape anisotropy would induce a perfect in-plane orientation of the magnetization and a homogeneous state with an external applied field. For the case of a large enough $K^{\perp}_{\rm S}>0$ for a thickness below a critical value ($d<d_{min}$) the magnetization would again be  homogeneous but with out-of-plane orientation and for $d>d_{max}$ an homogeneous in-plane state is expected. However, for a transition region $d_{min}<d<d_{max}$ different inhomogeneous states can be formed. Some of them can be modelled by a simple analytical model or by micromagnetic simulations as for instance in \cite{usov}. On the other limit case, for very large thickness, the situation is similar although the origin is different. For large thickness values, the demagnetizing field responsible for the in-plane orientation is weakened allowing for the formation of inhomogeneous states similar to the previous ones.

\begin{figure}[t]
    \includegraphics[width=0.8\columnwidth]{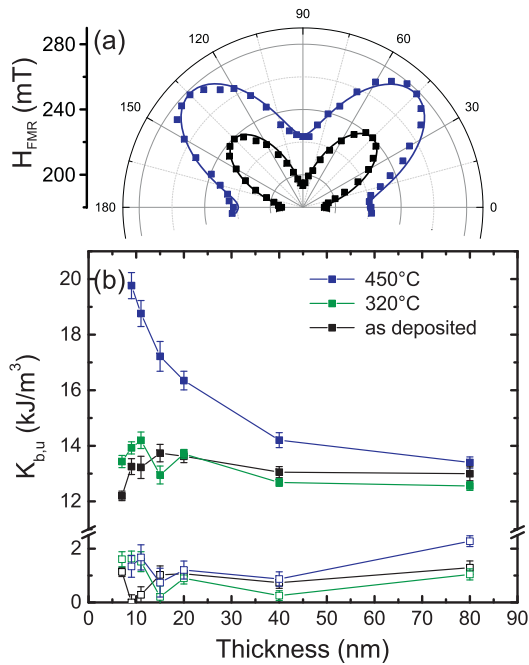}
	  \caption{\label{inplane-aniso}(Color online) (a) Dependence of $H_{\rm FMR}$ on the azimuthal angle $\varphi$ for 11~nm thick films for the as-deposited  and the 450$^{\rm o}$C annealed samples. The lines are a fit to Eq.~\ref{HFMRpolar}. (b) Dependence of the in-plane biaxial anisotropy constant $K_{\rm b}$ (filled points) and the in-plane uniaxial anisotropy constant $K_{\rm u}$ (hollow points) on the thickness $d$ for the as-deposited and the annealed series.}
\end{figure}

The in-plane anisotropies were studied by measuring the dependence of the resonant field $H_{\rm FMR}$ on the azimuthal angle $\phi$. Fig.~\ref{inplane-aniso}(a) shows exemplarily this dependence for a thickness of 11~nm in the range 0-180$^{\rm o}$ at 18~GHz  for the as-deposited sample and the 450$^{\rm o}$C annealed one. An overall four-fold anisotropy, as expected from the cubic lattice of CFA and the (100) growth direction is observed. The easy axes correspond to 0$^{\rm o}$ and 90$^{\rm o}$. Overimposed to this,  an additional weaker two-fold uniaxial anisotropy is also observed ( $H_{\rm FMR}$ at 0$^{\rm o}$ and 90$^{\rm o}$ are slighty different). The uniaxial anisotropy may be induced by stress in the film or by the vicinal structure in the substrate surface induced by miscut.

In order to extract the anisotropy fields the following formula was used:

\begin{equation} \label{HFMRpolar}
H_{\rm FMR}= \bar{H}_{\rm FMR}+ H_{\rm b}\text{cos}(4\phi)+H_{\rm u} \text{cos}(2\phi+\varphi)
\end{equation}

Here $H_{\rm b}$ and $H_{\rm u}$ are the biaxial and uniaxial anisotropy fields, $\phi$ is the in-plane azimuthal angle and $\bar{H}_{\rm FMR}$ is the averaged value. The angle $\varphi$ allows for a misalignment of the uniaxial and biaxial contributions, i.e. the easy axis of both contributions may be at different angles. The lines in Fig.~\ref{inplane-aniso}(a) are fits to this formula. These field values are related to the anisotropy constants $H_{\rm b,u}=\frac{2K_{\rm b,u}}{M_s}$.

The results for $K_{\rm b}$ and $K_{\rm u}$ from the fits are plotted in Fig.~\ref{inplane-aniso}(b). For the calculation of the anisotropy constant the magnetization values obtained from the fits in Fig.~\ref{ks-annealing} are used. For $K_{\rm b}$ we observe a  different  thickness dependence for the as-deposited series and the series annealed at 320$^{\rm o}$C compared to the series annealed at 450$^{\rm o}$C. The value of $K_{\rm b}$ shows minor variation for the as-deposited samples with a small reduction for the thinner films. The evolution is similar for the 320$^{\rm o}$C case. On the contrary, the anisotropy constant increases continously and strongly with decreasing thickness in the annealed series.  However, the values converge for thick films and for 80~nm the difference vanishes. This points to an important role of the stress in the films, which normally relaxes with thickness, in the evolution of $K_{\rm b}$.     The absolute values are in agreement with literature data \cite{belme2}.  The values of $K_{\rm u}$ are a order of magnitude smaller and the absolute values and the thickness dependence are very similar for the three cases. 

\section{Conclusions}

In summary, we measured the evolution of the interface induced perpendicular anisotropy for epitaxial CFA/MgO interfaces and we observed a strong increase with the annealing temperature up to a value of $K^{\perp}_{\rm S}=2.01\, \pm \,0.7$~mJ/m$^2$ for an annealing temperature of 450$^{\rm o}$C. A stabilization of a perpendicular magnetization orientation is then expected for films thinner than 1.7~nm. We studied the thickness dependent magnetic properties of CFA for as-deposited and annealed series. We obtained  minimum values for $\alpha$ for a thickness of 40~nm for all series and a different evolution with annealing for thinner or thicker films. We correlate this with interface and bulk changes upon annealing, respectively. The study of the in-plane anisotropy constant shows a much larger thickness dependence on the annealed samples compared to the as-deposited ones.

\section*{Acknowledgements}

Financial support by  M-era.Net through the HEUMEM project  is gratefully acknowledged.

\end{document}